# Health Behaviour Change Techniques in Diabetes Management Applications: A Systematic Review

**Ahmed Fadhil**
*University of Trento*
*Trento, Italy*
ahmed.fadhil@unitn.it

**Yunlong Wang**
*University of Konstanz*
*Konstanz, Germany*
yunlong.wang@uni-konstanz.de

**ABSTRACT**
The rapid growth in mobile healthcare technology could significantly help control chronic diseases, such as diabetes. This paper presents a systematic review to characterise type 1 & type 2 diabetes management applications available in Apple's iTunes store. We investigated "Health & Fitness" and "Medical" apps following a two-step filtering process (Selection and Analysis phases).
We firstly investigated the apps compliance to the persuasive system design (PSD) model. We then characterised the behaviour change techniques (BCTs) of top-ranked apps for diabetes management. Finally, we checked the apps regarding the stages of disease continuum. The findings revealed apps incorporation some PSD principles based on their configuration and behaviour change techniques. Most apps miss the element of BCT and focus on measuring exercise and caloric intake. Few apps consider managing specific diabetes type, which raises doubts about the effectiveness of those apps in providing sustainable diabetes management. Moreover, people may need multiple apps to initiate and maintain a healthy behaviour.

**INTRODUCTION**
Diabetes is one of the leading causes of death. It can lead to many life-threatening complications. In Europe around 60 million people, or 10.3% of men and 9.6% of women aged 25 years and over, are diagnosed with diabetes. According to the WHO[1], about 347 million adults in the world are diagnosed with diabetes, and deaths caused by diabetes will double between 2005 and 2030 [20]. There are three types of diabetes: type 1, type 2, and gestational diabetes. Type 1 diabetes is known as insulin-dependent diabetes (IDDM)[2], which often begins in childhood. Type 2 diabetes is known as non-insulin dependent diabetes (NIDDM)[3], which is more prevalent among adults. The prevalence of type 2 diabetes is largely due to an increase in poor lifestyle, including unhealthy dietary habits and physical inactivity. The gestational diabetes occurs when pregnant women with no pervious diabetes history develop a high blood glucose level. This type has an increasing risk of developing type 2 diabetes after pregnancy [20].
Studies have reported the substantial benefits of leading an active lifestyle, such as improved glycemic control and numerous physical, mental and social improvements. The rapid mHealth evolution could help change poor lifestyle, and thus promote health, prevent diabetes among healthy people, and help the treatment of diabetes patients [25], [15], [16].
In the present study, we systematically reviewed mobile apps available in the Apple iTunes store and evaluated their effect in diabetes management. The evaluation started with a search of type 1 or type 2 diabetes management apps. We conducted a descriptive analysis with a sample of 204 applications, mostly selected from the "Health & Fitness" and "Medical" categories of the iTunes store. After applying the inclusion and exclusion criteria, we classified apps based on their compliance with PSD model categories and their elements (e.g., Tailoring, Reminders and Surface Credibility) [24] and their incorporation of BCTs (e.g., Social support, Personalised feedback and Tailoring) [17]. Finally, we list out their focus of stage of disease continuum (e.g., health promotion or prevention from deterioration and complications) [4]. To our knowledge, this is the first systematic review that considered the characteristics and efficacy of mobile apps in terms of PSD and BCTs adaption to promote healthy behaviour and manage diabetes.

**RELATED WORK**
There is a strong correlation between diabetes and lifestyle related factors, such as physical activity, diet, weight, among others [12] that lead to diabetes as a health consequence. Developing techniques to efficiently manage diabetes has been the focus of many healthcare organisations and research institutes [7]. Despite of many available strategies in the field of diabetes management, to our knowledge, most techniques have shown only moderate effects to boost behaviour change and promote diabetes condition [7]. Those strategies didn't change poor dietary adherence nor give exercises recommendation among diabetes patients, which leads to further health complications. Adapting PSD and BCTs techniques have proven effective in mitigating the burden of poor lifestyle and its relation to diabetes [12].

---
[1] WorldHealthOrganisation:http://www.who.int/diabetes/en/

[2] American Diabetes Association: http://www.diabetes.org/diabetes-basics/type-1
[3] American Diabetes Association: http://www.diabetes.org/diabetes-basics/type-2

To induce users towards healthier lifestyle, a study by Bailoni et al., [3] developed PERKAPP, a context-aware system that uses a combination of persuasive technology, natural language generation and deep knowledge representation tools. The study found that personalised messages generated according to user preferences and the context can increase the persuasion effect in behaviour change. Studies investigated BCTs integration into product design to provide potential assistance in health-related prevention services, e.g., rewarding users for measuring caloric intake. Moreover, leveraging social influence can play a role in lifestyle promotion and prevention from escalation into chronic diseases including diabetes, which impacts motivation and behaviour change [18].

Mobile healthcare apps are increasingly used to promote healthier diet and physical activity. A study by West et al., [27] investigated the extent to which diet apps' content were guided by health behaviour theory in their design. The findings revealed that most apps were theory deficient. The study argued the need to incorporate health behaviour theories into individually tailored apps development. According to Freeland et al., [9], the overall dietary pattern is the most important factor of healthy eating. In contrast to the total diet approach, classification of specific food as good or bad is overly simplistic and can foster healthy eating behaviour. Achieving a healthy lifestyle requires an interdisciplinary approach between HCI and behavioural science. To bridge these two disciplines, we must discuss ways in which behavioural theory can inform research on behaviour change technologies [10]. Another study by Conroy et al., [5] investigated the BCTs adaption in mobile apps to promote physical activity. The findings revealed most apps description incorporated less than four BCTs. The most common techniques include providing instruction on how to perform exercise, providing feedback on performance, goal setting for PA and planning social support [5]. Studies [11,26] on diet and physical activity apps to promote healthy eating and energy expenditure revealed no clear evidence on their effectiveness to promote health and wellness. Yet, diabetes is strongly related to these lifestyle factors that have a significant effect on diabetes prevention and management. A study by Sama et al., [22] described the most common mobile health apps available on iTunes marketplace. The price, health factor and methods of engagement were the main outcome measures of the analysis. The findings revealed that most fitness and self-monitoring apps showed limited approaches, which calls for improvements. In the same context, a study by Cowan et al., [6] analysed 127 apps from the iTunes store in Health & Fitness category. The study established a theory-based instrument to rate apps inclusion of techniques from behavioural theories. The result revealed that "health belief model[4]" was the most prevalent theory, however apps contained only minimal theoretical background. The study suggested a collaboration between health behaviour change experts and app developers to foster apps superior both in theory and programming, and hence result in better mHealth approach.

Integrating persuasive technologies is found to encourage physical activity. A study by Munson et al., [19] designed an app with goal-setting technique and deployed it into four-week field study. The findings revealed participant's acceptance to have primary and secondary goals and receiving non-judgmental reminders. However, trophies and rewards failed to motivate most participants. This raises questions about reward design and application [19]. Persuasive messages and reminders are widely disseminated and utilised in individual adherence to regular physical activity regime. While research has examined strategies for physical activity messages, according to Latimer et al., [13] there has been no systematic effort to examine optimal message content. The study investigated three approaches to construct physical activity messages, namely message tailoring, message framing and targeting messages to affect change in self-efficacy [13]. Overall, the study provided strong evidence and recommendations for optimal message content and structure. In a study of mobile system for diabetes management [14] the participant had to create more regular meal pattern and increase exercise and regularity of their blood glucose level. In a different study by Ristau et al., [21] the effect of mobile apps was measured and found that it led to statistically significant improvement in A1C[5] in adults with type 2 diabetes. The study concluded that considering patients need (e.g., their age, app cost and specific (HCPs) factors to consider when recommending a self-management diabetes application) by healthcare providers is vital to maximise the benefits of the app.

Although the increasing prevalence of mobile apps in weight management, there is few evidences about integrating behaviour change theories into their design. A work by Azar et al., [1] evaluated diet and anthropometric measurements tracking apps based on incorporated features consistent with behaviour change theories. The results divided apps into five categories, ranging from diet tracking, healthy cooking, weight tracking, grocery and restaurant decision making. The findings demonstrated low adaption of BCTs and persuasive technology in app design and implementation [1]. Another study by Bailoni et al., [2] used a diet-oriented concept to classify foods based on their composition and the dressing to ease user monitoring activities. Hence, the aim was to map common-sense dishes, such as "Pasta alla Carbonara" [2] to the set of basic foods with their nutritional information.

## METHOD

### Aims

This review investigates mobile application in type 1 & type 2 diabetes management, then identifies the features in highly rated apps based on the PSD, BCTs and the stages of disease continuum. This will help identify gaps and guide future research directions.

### Search Strategy

A systematic search of diabetes management apps in iTunes store was conducted in September 20, 2016. The search terms

---
[4] http://www.jblearning.com/samples/0763743836/chapter%204.pdf

[5] https://medlineplus.gov/a1c.html

were structured as Term:
"Diabetes" || "Diabetes Mellitus" || "Diabetes type 1" || "Diabetes type 2" || "Blood Sugar" || "Insulin" || "Glucose" || "A1c". The search returned 522 results which were filtered following a screening process. The apps acquired from "HEALTH & FITNESS", "MEDICAL", "UTILITY", "SOCIAL NETWORKING", "EDUCATION", "INSTRUCTION", "REFERENCE", "FOOD & DRINK", "PHOTO & VIDEO", "BOOKS", AND "PRODUCTIVITY" categories.

Only freely available apps were included in the study and there were no language restrictions, although almost all apps met in- clusion criteria are in English. The description about each app was obtained from either the iTunes store or from their official website. Data such as, platform support, app category, custom ratings, license type, app ratings, and last updated date were all obtained from the iTunes store, although some had missing values. For example, some apps were lack of custom ratings and therefore excluded from the study.

**Data Sources**
The total number of apps after the screening process (see Figure-1) was 204 apps. The search criteria terms and the corresponding number of apps together with their categories are listed in Table-1.

**Table 1.** Search Terms, # Applications and Categories.

| Terms | # Apps | Category |
|---|---|---|
| Diabetes | n = 86 | Health & Fitness, Medical, Social Networking, Reference, Books, Food & Drink, Photo & Video, Productivity |
| Diabetes Mellitus | n = 20 | Health & Fitness, Medical, Instruction |
| Diabetes type 1 | n = 4 | Health & Fitness, Medical |
| Diabetes type 2 | n = 1 | none |
| Blood Sugar | n = 38 | Health & Fitness, Medical |
| Insulin | n = 25 | Health & Fitness, Medical, Education |
| Glucose | n = 19 | Health & Fitness, Medical, Utility, Social Networking |
| A1c | n = 11 | Health & Fitness, Medical |

**Application Selection**
The selection condition included apps addressing health issues that correlate to diabetes. The sample apps obtained on November 12, 2016 included one or more health factors (e.g., diet, physical activity and diabetes management).

**Inclusion & Exclusion Criteria**
This was achieved with two-step process (namely, Selection and Analysis phase) by manually reviewing each app by the coder and checking for Customer Ratings, License, App Ratings and Date Updated. The rated apps inclusion condition was: ($\forall$ App Ratings(Customer Ratings $\geq$ 4 & License $\in$ Free & Updated Date $\geq$ 2014)) (see Table-2 for the Selection Phase). According to the inclusion criteria, 22 apps were obtained, mostly belonging to HEALTH & FITNESS and MEDICAL categories, with two apps belonging to BOOKS and FOOD & DRINK categories.

**Table 2.** Selection Phase and Inclusion Criteria.

| Selection Phase | Condition |
|---|---|
| App Ratings | $\forall$ |
| Customer Ratings | $\geq 4$ |
| License | $\in$ Free |
| Date Updated | $\geq 2014$ |

The Analysis Phase checked for apps input/ output mechanism, monitoring the condition, coaching, social influence, and applied gamification. We list and define in Table-3 the analysis phase categories and provide an example per category. This phase excluded three apps since they were focused on other domains and had no connection with diabetes management. Hence, the final list was 19 apps eligible for the analysis.

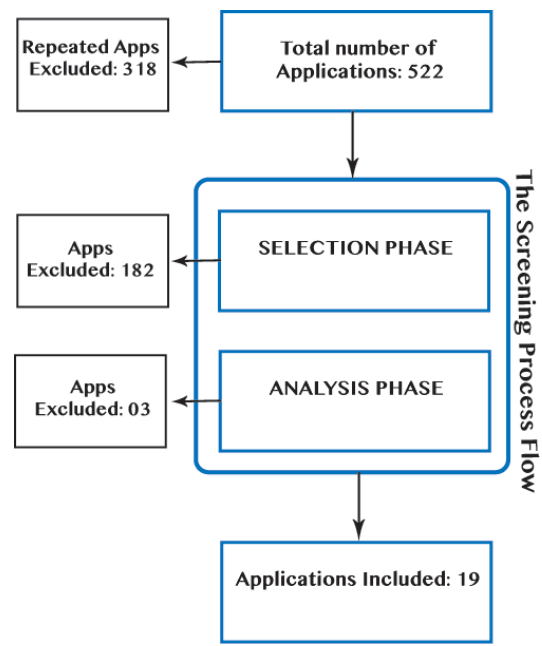

**Figure 1.** The Screening Process for Diabetes Apps.

**Diabetes Application**
Diabetes apps ranged from personal, professional, or general information guides and reference about diabetes. Majority of apps were either for type 1 or type 2 diabetes to measure daily blood sugar, to perform fitness and physical activities, nutrition and dietary support, reminder about medication dosages; or for caregivers and healthcare professionals to support the diagnoses and treatment of diabetes.

**FINDINGS**

**Systematic Review**
The two-step processes provided insights on techniques and missing elements across the apps. Based on the coders review, most apps incorporated no BCTs. Moreover, the apps focus on health promotion and prevention aspects of diabetes. The majority of apps were focused on diet, physical activity promotion and weight management. We analyzed whether an app provides hybrid, human expert, or hybrid + human expert feedback. We particularly focused on the support type and the stage at which the support is given (before, during or after the activity fulfilled). We found that majority of apps provided the support only after the activity fulfillment (by both hybrid and human expert). However, there was no human support before and during the activity performance and was limited to support users after the activity has been fulfilled (see Figure-2). The review provided insights into the evolution of mobile healthcare apps and the opportunities to improve status of end users.

Table 3. Analysis Phase Category with Definition and Example.

| Analysis Phase Category | Definition | Example |
|---|---|---|
| App input/ output mechanism | The interaction type between the user and the system, their medium, the input and output to and from the system. | The system promotes user to provide their daily steps and dietary adherence, or blood glucose measure as input, then it provides them with diabetes adherence data |
| Condition Monitoring | The way the system lets user track/manage their activities and the type of monitoring support it provides. | Self-monitoring or with coach support monitoring of users overall diabetes management and adherence |
| Coaching | The type by whom the support is given, whether its hybrid or human expert based. This also involves the interaction phase at which the support is given to the user. | A coaching by a human expert to track and adhere user to a healthier lifestyle before, during or after the activity accomplishment |
| Social Influence | The social support provided by external actors outside the system. This could include, healthcare provider, family or friends. | Family members track users' glucose measurement or medication adherence and provide motivational feedbacks to the user |
| Applied Gamification | The strategy or type of motivational elements integrated within the application to provide the necessary engagement and support to the user. | Motivating user to follow an active lifestyle by rewarding them to the goal achievement |

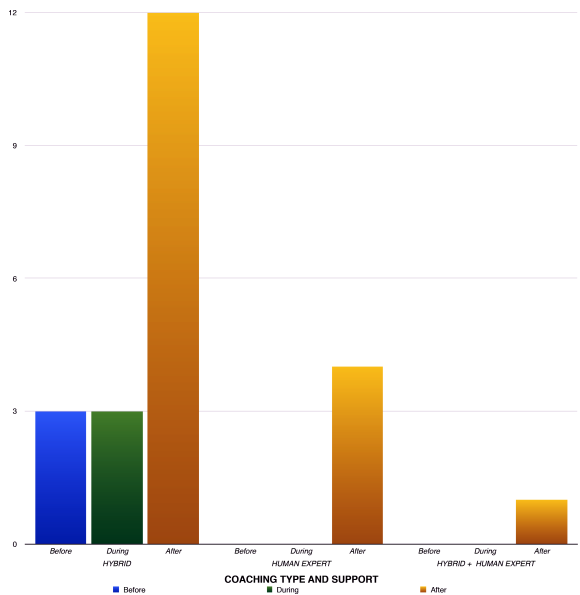

Figure 2. The Coaching Type & Support Stage.

**Persuasive System Design (PSD) Analysis**
This model is applicable to systems designed to form, alter or reinforce attitude, behaviour, or a compliance act without deception, coercion or inducements. Although the PSD shows some overlap with BCTs, since both aim to influence our mind or behaviour. However, the PSD model is focused on changing behaviour by means of technology. The input from PSD and BCTs can be combined when designing mobile health applications, as they complement each other.

We applied the PSD model to the obtained apps and checked for their compliance with the PSD principles (Primary Task Support, Dialogue Support, System Credibility Support, and Social Support) [24] and their various persuasive elements (the PSD model, its categories and accompanying elements appear in Table-4). The principles are the essence to design a persuasive technology [24]. Whereas the elements are based on what technology can do to persuade its users into changing their attitude or behaviour. Based on the analysis, Personalisation (e.g., provide personalised information and most frequent feature) was the mostly applied element among all the analysed apps (Frequency = 14 (11.7%)). Followed by Tailoring (e.g., tailor the activity to user preferences) and Self-monitoring (e.g., allow user to track and monitor own health) (Frequency = 13 (10.8%)). Then, Expertise (Frequency = 11 (9.17%)), Trustworthiness (Frequency = 9 (7.5%)), Reminders (Frequency = 8 (6.67%)), Cooperation and Normative influence (Frequency = 7 (5.83%)), Third-party endorsements, Social learning (Frequency = 4 (3.45%)), Reduction, Tunnelling, Suggestion, Social comparison, Social facilitation and Recognition (Frequency = 3 (2.59%)), Surface credibility and Competition (Frequency = 2 (1.72%)).

However, Simulation, Rehearsal, Praise, Rewards, and Authority were the least applied principles among all the analyzed apps (Frequency = 1 (0.87%)). Finally, Similarity, Liking, Social role, Real-world feel, and Verifiability were missing in all the apps (Frequency = 0 (0%)).

Reflecting on the four categories; primary task support and system credibility support were more frequently applied by most apps, while social support and dialogue support appeared less frequently.

**Behaviour Change Techniques (BCTs) Analysis**
Although the abundant literature approaches to behaviour change, there was a minimal incorporation of BCTs into app design. Collecting BCT related data could guide developers to refine their offering for improved impact. Table-5 highlights the various behaviour change techniques and their frequency in the apps.

**Table 4. The PSD Principles & Persuasive Elements with Frequency.**

| Primary Task Support | Dialogue Support | System Credibility Support | Social Support |
|---|---|---|---|
| Reduction(1.1)= 5 | Praise (2.1)=1 | Trustworthiness (3.1) =9 | Social learning (4.1)=3 |
| Tunnelling (1.2)=3 | Rewards (2.2)=1 | Expertise (3.2) =11 | Social comparison (4.2)= 4 |
| Tailoring (1.3)=13 | Reminders (2.3)= 8 | Surface credibility (3.3)=2 | Normative influence (4.3)= 7 |
| Personalisation (1.4)= 14 | Suggestion (2.4)=3 | Real-world feel (3.4)=0 | Social facilitation (4.4)=3 |
| Self-monitoring (1.5)=13 | Similarity (2.5)=0 | Authority (3.5)= 1 | Cooperation (4.5)=7 |
| Simulation (1.6)= 1 | Liking (2.6)= 0 | Third-party endorsements (3.6)=5 | Competition (4.6)=2 |
| Rehearsal (1.7)= 1 | Social role (2.7)= 0 | Verifiability (3.7)= 0 | Recognition (4.7)= 3 |

**Table 5. Behaviour Change Techniques Applied in Applications.**

| Frequency Apps | Behaviour Change Techniques (BCTs) |
|---|---|
| n=8 | Social support |
| n=1 | Personalised feedback |
| n=5 | Self-monitoring |
| n=4 | Reminders |
| n=5 | Tailoring |
| n=1 | Reinforcement |
| n=1 | Goal-setting |
| n=1 | Challenge |
| n=1 | Comparison |
| n=1 | Collaboration |

**Stages of Disease Continuum**

During the analysis phase, we investigated the application position at the stages of disease continuum (see Table-6), which refers to the stage at which the app service.

**Table 6. The Stages of Disease Continuum.**

| Stages of Disease Continuum | Frequency Apps |
|---|---|
| Health Promotion | n=7 |
| Prevention From Complications | n=5 |
| Prevention From At-risk | n=2 |
| Disease Prevention | n=1 |
| Prevention From At-risk, Disease Prevention | n=1 |
| Health Promotion, Prevention From Complications | n=1 |
| Health Promotion, Prevention From At-risk | n=1 |
| Health Promotion, Disease Prevention | n=1 |

**DISCUSSION**

Behavior change experts could work with app developers to incorporate BCTs into app development tailored for health promotion [27]. More work needs to consider the Dialogue Support and Social Support in the PSD model. Moreover, diabetes is strongly correlated to diet, hence providing patients with healthy food choices is essential to increase their knowledge about food and understand the barriers for healthy food choices. More work is needed to deeply analyze BCTs application in app design and implementation, since they form the foundation for app effectiveness in tackling various health issues.

There is a mixed picture on the effectiveness of persuasive technology and gamification in the context of diabetes. Patients are heterogenous (e.g., they have different level or preparedness and different needs). Gamification in diabetes apps impacts patients at different levels by providing motivation and engagement which varies among individuals [23]. Few apps considered patients need, cultural attitude, age, gender or geolocation. We believe applying game elements for routine self-management tasks could increase app sustainability.

**Limitations**

This study focused on diabetes apps on iTunes store for iOS. Android apps were excluded due to the huge number of apps and time limitation, and the common apps within both iTunes and Google Play stores. Moreover, only freely available apps were considered in the review, since they're reached in big portion by the population, which was sufficient for the analysis. In addition, privacy and security aspects, user friendly GUI or user experience design were excluded from the evaluation since they fall outside the scope. Finally, our investigation revealed most apps with human expert support were focused on providing the support after the activity execution. This question the effectiveness of the intervention, since providing instance support is important to guide patients in their journey and provide them with emotional support, and hence for the success of the application [8].

The positive outcome of the analysis was the primary task support and social influence (e.g., family and friends support) which were present in the majority of the apps. This enhances user stimulation to engage in their health promotion. Application developers should consider patients heterogeneity and target a group of patients with similar preferences, and hence provide a more tailored support.

**CONCLUSION**

Engaging users in their healthcare is a hard task, especially when it's tied to behavior change. This study performed a systematic review on a range of type 1 & type 2 diabetes apps. Based on the analysis, the majority of apps focused on diet and physical activity. We identified apps compliance with the PSD model and adaption of BCTs. The findings revealed limited adaption of BCTs by app developers. This presents an opportunity to improve the effectiveness of such technology. Moreover, drawing users into specific apps could have a significant impact on their health outcomes. We need to have information about user preferences, ability and most importantly emotion. Finally, adding motivational techniques, such as game elements could help to measure how realistic is the concept and how far it can go with respect to user attitude change. To conclude, there is an abundant space for development in diabetes apps, and future direction should map BCTs and PSD model to support both the theory and technology in diabetes app design and implementation.